\documentstyle[floats,multicol,aps,epsf,prl]{revtex}

\begin{document}
\draft
\twocolumn[\hsize\textwidth\columnwidth\hsize\csname @twocolumnfalse\endcsname
\title{
High field study of normal state magneto-transport in 
$\bf \rm Tl_2Ba_2CuO_{6+\delta}$
}
\author{A. W. Tyler$^1$\cite{Aapm}, 
Yoichi Ando$^2$\cite{Aando}, F. F. Balakirev$^2$,
A. Passner$^2$, G. S. Boebinger$^2$,\\
A. J. Schofield$^3$, A. P. Mackenzie$^1$\cite{Aapm} and O. Laborde$^4$}
\address{$^1$IRC in Superconductivity, Madingley Road, Cambridge, CB3 0HE,
United Kingdom}
\address{$^2$ Bell Laboratories, Lucent Technologies, 700 Mountain Avenue,
Murray Hill, New Jersey 07974, U. S. A.}
\address{$^3$T.C.M., Cavendish Laboratory, Madingley Road, Cambridge, CB3 0HE,
United Kingdom}
\address{$^4$ Centre de Recherche sur les Tr\`es Basses Temp\'eratures,
Laboratoire des Champs Magn\'etiques Intenses, C.N.R.S. BP 166, 38042
Grenoble-c\'edex, France}
\date{\today}
\maketitle
\begin{abstract}
We present a study of in-plane normal state magneto-transport in single
crystal $\rm Tl_2Ba_2CuO_{6+\delta}$ in 60T pulsed magnetic fields.  In
optimally doped samples (${T_c \sim 80K}$) the weak-magnetic-field regime
extends to fields as high as 60T, but in overdoped samples (${T_c \sim 30K}$)
we are able to leave the weak field regime, 
as shown by the behavior of both the
magnetoresistance and the Hall resistance.  Data from samples of both
dopings provide constraints on the class of
model necessary to describe normal state transport in the cuprates.
\end{abstract}
\pacs{PACS numbers: 72.15.Gd, 74.72.Fq, 72.10.Bg}
\vskip1pc
]

Measurements of the electrical transport properties of the high-$T_c$
cuprate superconductors have proved to be one of the most useful
ways of probing their anomalous normal state properties.  Studies of
the Hall effect in $\rm YBa_2Cu_3O_{7-\delta}$\cite{chien} show that near
optimum doping,
the resistivity ($\rho$) has an approximately linear temperature
dependence, while the inverse Hall angle ($\cot \theta_H$) varies
nearly quadratically with temperature.  Systematic measurements of
these properties have been made in many cuprate materials,
as a function of disorder and carrier concentration.  Although some
exceptions have been reported, the experimental picture which has
emerged is that $\cot \theta_H$ continues to vary approximately
quadratically in temperature even when $\rho$ changes its temperature
dependence from linear on underdoping or overdoping.

Most attempts to understand this intriguing behavior fall into two
broad classes.  In the first (the ``two-lifetime'' models), the
experimental observations are taken to indicate the existence of two
separate and essentially decoupled scattering times governing
longitudinal transport and Hall transport, often referred to as
$\tau_{tr}$ and $\tau_H$ respectively\cite{pwa,cst}.  These two
intrinsic lifetimes are assumed to co-exist at all points on the Fermi
surface (FS).  Theories constructed from this starting point are
intrinsically non-Fermi liquid in nature.  The more conventional class
of models (the ``anisotropic'' models) exploits the four-fold
anisotropy of the cuprate Fermi surfaces and postulates the existence
of an anisotropic scattering rate whose magnitude and temperature
dependence is different on different parts of the
FS\cite{carrington,kendziora,pines}.

To date, almost all normal state magnetotransport studies on the
cuprates have been performed in magnetic fields of less than 20T,
in the weak-field regime ($\omega_c \tau \ll 1$,
where $\omega_c$ is the cyclotron frequency and $\tau^{-1}$ is the
scattering rate).  In this regime, the magnetoresistance is parabolic
in $B$\cite{harris,kimura,hussey,balakirev}, and the Hall voltage is linear
in $B$.  Recently, Harris {\it et al.}\cite{harris} pointed out
that the magnitude of the weak-field (``WF'')
orbital normal state MR ($\Delta\rho/\rho^{WF}$) is proportional to
the variation of the local Hall angle around the Fermi surface.
Observations on $\rm YBa_2Cu_3O_{7-\delta}$ and optimally doped $\rm
La_{2-x}Sr_xCuO_4$ have shown that the
temperature dependence of the MR agrees with that of the square of the
Hall angle, favoring two-lifetime models for which $\tau_H$ exists at
all points on the FS\cite{harris}.

Here, we report magnetotransport results on single
crystals of $\rm Tl_2Ba_2CuO_{6+\delta}$ in 60T pulsed magnetic fields.  In
samples near
optimum doping ($T_c\sim 80K$), the weak-field regime is seen to
extend to nearly 60T, and we see behavior similar to that observed in
$\rm YBa_2Cu_3O_{7-\delta}$ and optimally doped $\rm La_{2-x}Sr_xCuO_4$.  In
strongly overdoped samples
($T_c\sim 30K$), however, we are able to observe significant
deviations from the weak field forms of $\rho_{xx}(B)$ and
$\rho_{xy}(B)$. We find that the high field magnetotransport provides
an independent way of estimating an important parameter,  $\omega_c \tau$,
and thus provides additional constraints to the existing transport 
models.  Analysis of
our data favors the general class of two-lifetime models to
describe normal state transport in the cuprates.

\begin{figure}[tb]
\epsfxsize=3.5in
\centerline{\epsfbox{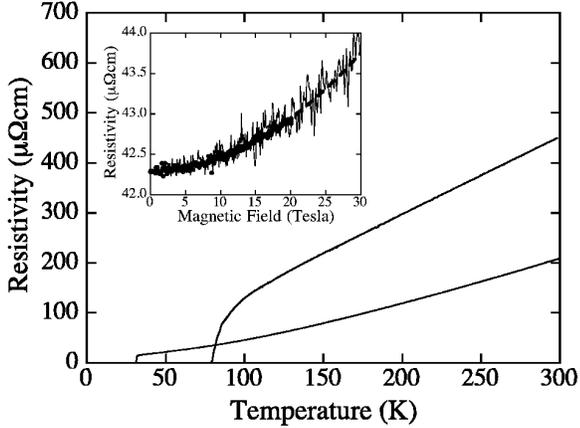}}
\protect\caption[*]{The resistivity in zero magnetic field of optimally 
doped ($T_c\sim 80K$) and overdoped ($T_c\sim 30K$) single crystals of
$\rm Tl_2Ba_2CuO_{6+\delta}$.  
The inset shows a comparison of magnetoresistance (MR)
measurements in d.c. (circles) and pulsed (solid line) magnetic fields
at $T\sim 115K$ for an overdoped sample.  The MR in this range can be
well fitted with a quadratic field dependence (dashed line).  The
pulsed field MR data were multiplied by a factor of 1.015 to correct
for a small difference in the temperature at which the pulsed and
d.c. measurements were made.
\label{fig1}}
\end{figure}

Tetragonal single crystals of $\rm Tl_2Ba_2CuO_{6+\delta}$ were grown
using a Cu-rich flux. With $\rm Tl_2Ba_2CuO_{6+\delta}$, 
degradation of the surface quality
on annealing is a persistent problem, leading to poor quality
electrical contacts for the transport measurements.  In our
experience, the best way of making contacts to samples with elevated
$T_c$ is to evaporate gold contact pads onto the as-grown sample.  The
desired $T_c$ values were then achieved by annealing using various
combinations of annealing temperature, atmosphere and time.  The
anneal also diffuses the gold into the sample, providing a
low-resistance contact to which gold wires are attached using a silver
paint (Dupont 4929) which hardens in air at room temperature.  Care
was taken when evaporating the gold to ensure that it covered the
sides of the sample.   The resistivity for optimal ($T_c\sim 80K$) and
overdoped ($T_c\sim 30K$) samples is shown in Fig.~\ref{fig1}.

A resistive d.c. magnet in Grenoble was used for MR measurements in
fields up to 20 tesla, using a standard four terminal
a.c. technique with a 32 Hz excitation frequency.  Temperature
stability of 1 part in $10^4$ was achieved using a capacitance sensor
which is insensitive to field.  The temperature was measured in zero
field before and after each sweep, to check for possible temperature
drift.  The longitudinal MR was also measured at several temperatures
and found to be at most 5\% of the total MR.  The total measured
transverse MR is thus dominated by the orbital MR for all of the data
in the figures.

Precise measurement of normal state magnetotransport in a pulsed field
presented a considerable challenge, due to the higher noise levels and
the relatively low resistances of the $\rm Tl_2Ba_2CuO_{6+\delta}$
crystals (as little as $10m\Omega$ below $50K$).  After some
experimentation, voltage sensitivity of better than 1 $\mu$V could be
achieved at excitation currents of 5-10mA.  The temperature was
stabilized and measured in zero field. Comparison of up- and
down-sweeps was used to check for eddy current heating, which was
largely avoided due to the very small sample sizes (typically $300
\times 100 \times 15\mu m^3$).  Also, all the data shown in this paper
are at sufficiently high temperatures that the MR is dominated by
normal state orbital contributions rather than superconducting
fluctuations.

Given the difficulty of using a pulsed field to measure small MR
effects in low resistance samples, we first established, using four
different samples, that the results using d.c. and pulsed techniques
are consistent.  As an example, in the inset to Fig.~\ref{fig1} we
show the resistivity at $115K$ of one of the overdoped samples
measured as a function of magnetic field up to 20T in the resistive
d.c. field (circles).  The data measured in the pulsed magnetic field
at Bell Labs (solid line) overlays the d.c. field data very well.

\begin{figure}[tb]
\epsfxsize=3.5in
\centerline{\epsfbox{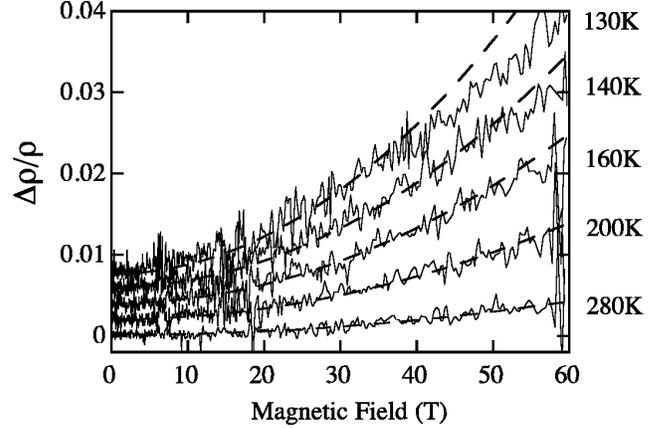}}
\protect\caption[*]{The MR of an optimally doped 
crystal of $\rm Tl_2Ba_2CuO_{6+\delta}$ 
($T_c\sim 80K$) at a series of temperatures between $130K$ and $280K$.
Data at different temperatures have been offset for ease of viewing.
The dashed lines are parabolic fits to the data.
\label{fig2}}
\end{figure}

Fig.~\ref{fig2} contains pulsed field MR data from a sample near
optimal doping ($T_c\sim 80K$).  For this sample, the MR is small
(only 3\% at 60T and $130K$) and the field dependence is basically
quadratic for $T >130K$ (dashed line in the figure) over the whole
field range.  This is good evidence that the weak field regime is
applicable in fields as high as 60T in optimally doped $\rm
Tl_2Ba_2CuO_{6+\delta}$.

Since a single isotropic band has zero MR, the parabolic MR in
Fig.~\ref{fig2} might result from anisotropy, whether the scattering
rate varies around the FS\cite{carrington,kendziora,pines} or the
``local cyclotron effective mass'' $m^*$ (proportional to $1/v_F$)
varies due to the local geometry of the FS\cite{harris}.
Interestingly, it is also possible to construct two-lifetime models in
which neither $\tau$ nor $m^*$ are anisotropic, yet there is a
non-zero MR.  In one such model\cite{cst}, fast and slow relaxation
rates exist at all $\vec{k}$ points.  In terms of the MR, this is roughly
analogous to a two fluid picture, and a quadratic weak-field MR
results even for a parabolic band.  Thus, the model-specific part of
the data in Fig.~\ref{fig2} is not the parabolicity,
$\Delta\rho/\rho^{WF} =\alpha B^2$, rather it is the interpretation of
the prefactor $\alpha$ in terms of the lifetimes (or masses).

At optimal doping, the Hall effect in $\rm Tl_2Ba_2CuO_{6+\delta}$ gives a
very clear example
of two-lifetime behavior, with $\rho$ and $\cot \theta_H$ following
linear and quadratic temperature dependences to an accuracy that poses
a challenge to any theory of the cuprate normal state\cite{tyler}.
Harris et al.\cite{harris} have shown that in $\rm YBa_2Cu_3O_{7-\delta}$
and $\rm La_{2-x}Sr_xCuO_4$ near
optimal doping, the weak field MR obeys a special form of Kohler's
rule in which $(\cot \theta_H)^2 \Delta\rho/\rho$ scale at different
temperatures, while the traditional Kohler scaling
$\rho\Delta\rho/B^2$ is violated.  This is also the case for the MR
data from the optimally doped $\rm Tl_2Ba_2CuO_{6+\delta}$ of
Fig.~\ref{fig2}.  As discussed
in Ref.~\onlinecite{harris} and above, this observation favors
two-lifetime models in which lifetimes with two different temperature
dependences exist at all points of the FS.

\begin{figure}[tb]
\epsfxsize=3.5in
\centerline{\epsfbox{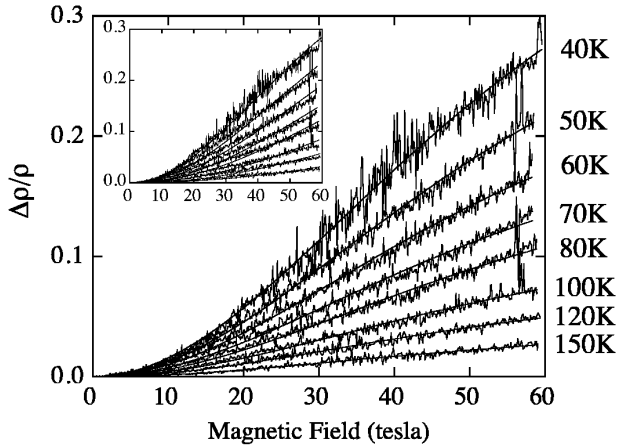}}
\protect\caption[*]{The MR of an overdoped crystal of 
$\rm Tl_2Ba_2CuO_{6+\delta}$ ($T_c\sim 30K$) at 
a series of temperatures between $40K$ and
$150K$.  The fits (solid lines) are produced by a general expression
for orbital MR beyond the weak field limit (Eq.~\ref{general}).  The
inset shows the same data fitted to Eq.~\ref{cstmr} with $\omega_c
\tau_{tr}$ fixed by the resistivity data of Fig.~\ref{fig1}.
\label{fig3}}
\end{figure}

Because the 60T magnetic field reveals a strictly parabolic MR in the
optimally doped sample, we turn our attention to overdoped samples
with lower resistivity to better reach the high magnetic field regime.
The MR of the overdoped samples is much larger, and deviates
significantly from a quadratic field dependence at high fields and low
temperatures.  Going beyond the weak field limit in any of the models
discussed above leads to a modified general expression of the form
\begin{equation}
{\Delta\rho \over \rho} = {\alpha B^2 \over 1 + \left(\beta B\right)^2}
\label{general}
\end{equation}
where $\beta$ is a second constant, principally related to the overall
scale of $\omega_c \tau$.  This expression accounts for the data in
Fig.~\ref{fig3} quite well, as shown by the curve fits, so our
observations are consistent with the onset of MR saturation as we
leave the weak field regime\cite{com1}.

To determine whether the high field MR can constrain which general
class of models is applicable to the cuprates, we examine the physical
origin of weak- and strong-field MR in more detail.  As discussed
above, in conventional models, the weak field MR is a measure of
anisotropy around the FS, although its magnitude is not easy to
interpret.  However, the onset of saturation is much easier to
interpret, because it arises once the quasiparticles orbit a
significant fraction of the Fermi surface between scattering events.
Thus $\beta \sim e\langle m^*/\tau\rangle^{-1}$, where $\tau$ is the
scattering lifetime and $\langle \rangle$ denotes a FS average.  The
product $\beta B$ gives the value of $\omega_c \tau$ determined from
the high field MR.

The different classes of model for the cuprate normal state make very
different predictions for the relationship between this high field
value of $\omega_c \tau$ and those that can be estimated from the
resistivity and weak field Hall effect.  If the Hall effect and
resistivity are to be understood in terms of an anisotropic
model\cite{carrington,kendziora,pines}, the fast and slow relaxation
rates which lead to $\tau_{tr}$ and $\tau_H$ exist on different parts
of the FS\cite{com2}.  When $\omega_c \tau_{tr}$ is estimated from
$\rho$ and $\omega_c \tau_H$ is estimated from $\theta_H$, we find
that $\omega_c \tau_H > \omega_c \tau_{tr}$ for all temperatures below
room temperature.  Since high field MR saturation would be expected to
be dominated by those regions with the highest scattering rate, the
value of $\omega_c \tau$ estimated from the high field MR would be
expected to be close to $\omega_c \tau_{tr}$ (and not to $\omega_c
\tau_H$).  We shall show that this is apparently not the case for our
data.

The opposite conclusion, that the high field MR is determined by
$\omega_c \tau_H$, is reached for two-lifetime models, two of which we
shall consider here.  In the picture discussed in
Refs.~\onlinecite{chien,pwa} and~\onlinecite{harris}, $\tau_H$ governs
all responses to a magnetic field.  Since $\tau_H$ is constant around
the FS in this model, some $v_F$ anisotropy is necessary for the
appearance of a non-zero MR.  Models of the FS of most cuprates show
that $v_F$ is smallest in the regions of highest FS curvature, so the
high field MR would be expected to give $\omega_c \tau$ with the
temperature dependence of $\tau_H$ and a magnitude which is determined
primarily by the curved parts of the FS.  Since a weak field
measurement of the Hall angle contains an even stronger weighting to
areas of large FS curvature\cite{ong}, the value of $\omega_c \tau$
obtained from high-field measurements would be expected to be similar
to that determined from weak field Hall measurements at all
temperatures.

In the formulation of a two-lifetime model described in Ref.
\onlinecite{cst}, a simple
analytical expression can be derived for the MR in terms of the
quantities $\omega_c \tau_{tr}$ and $\omega_c \tau_H$, if FS anisotropy is
neglected:
\begin{equation}
{\Delta \rho \over \rho} = {\omega_c \tau_H \left( \omega_c \tau_H -
\omega_c \tau_{tr} \right) \over \left( 1 + \omega_c^2 \tau_H^2
\right)} \; .
\label{cstmr}
\end{equation}

In both two-lifetime models, then, high-field saturation is expected
to probe $\omega_c \tau_H$.  Our data support this expectation.  The
value of $\beta$ fitted from the data at $40K$ where the saturation is
largest corresponds to $\omega_c \tau$ of 0.9 at 60T.  At that
temperature the weak field Hall angle per tesla is 0.01 $\rm T^{-1}$ 
giving an estimated value for $\omega_c \tau_H$ of 0.6 at 60T.
The measured $\rho$ can be used to estimate
$\omega_c \tau_{tr}$ for a large FS (corresponding to $k_F\sim
0.7$\AA$^{-1}$~\cite{mackenzie}) using the expression $\omega_c
\tau_{tr} = 2\pi d B/ek_F^2\rho$, where $d$ is the interplane spacing
of $11.6$\AA.  At 60T, $\omega_c \tau_{tr}$ is only 0.3, 
so the MR value for $\omega_c \tau$ derived from high fields 
certainly seems to be in better agreement with $\omega_c \tau_H$.

\begin{figure}[tb]
\epsfxsize=3.5in
\centerline{\epsfbox{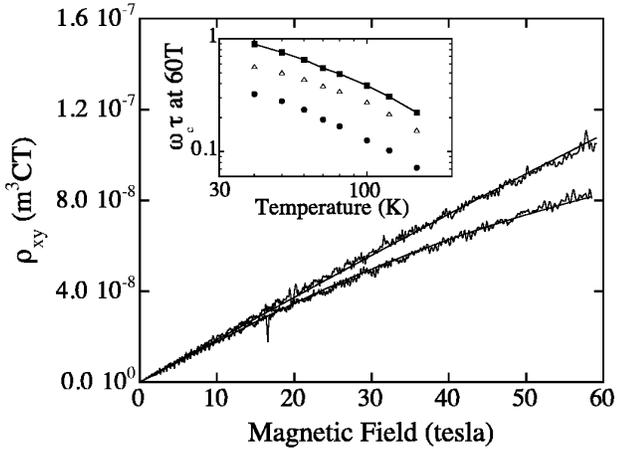}}
\protect\caption[*]{The Hall resistivity $\rho_{xy}$ 
for the $T_c\sim 30K$ 
crystal at $50K$ and $150K$.  The fits (solid lines) to
Eq.~\ref{csthall} were made with both $\omega_c \tau_{tr}$ and
$\omega_c \tau_H$ fixed by the fits to the MR data in Fig.~\ref{fig3}.
The inset shows the value of $\omega_c \tau_H$ obtained using this
procedure (squares joined by solid line) compared with measurements of
$\omega_c \tau_H$ from the weak-field Hall angle (triangles) and
$\omega_c \tau_{tr}$ from the resistivity (circles).
\label{fig4}}
\end{figure}

In the model of Ref.~\onlinecite{cst}, it is also possible to derive
an analytical expression for the Hall resistivity
\begin{equation}
\rho_{xy} = {m^* \omega_c \over n e^2} {\omega_c \tau_H \over \omega_c
\tau_{tr}} {\left(1 + \omega_c^2 \tau_H \tau_{tr} \right) \over
\left( 1 + \omega_c^2 \tau_H^2 \right)} \; ,
\label{csthall}
\end{equation}
where $n$ is the carrier concentration.  If we neglect the mild $v_F$
anisotropy round the FS that probably exists, we can reduce
Eq.~\ref{cstmr} to a single parameter fit for $\omega_c \tau_H$ by
estimating $\omega_c \tau_{tr}$ at each temperature from the
resistivity data of Fig.~\ref{fig1}.  The resulting fits are shown in
the inset of Fig.~\ref{fig3}. Using these values for $\omega_c
\tau_H$, fits using Eq.~\ref{csthall} are reduced to a single free
parameter: an overall scale factor related to $n$.
The $\rho_{xy}$ data of Fig.~\ref{fig4} are fitted
using Eq.~\ref{csthall} with both $\omega_c \tau_{tr}$ and $\omega_c
\tau_{H}$ fixed. The constrained fits to the MR data
are poorer than those in the main
panel of Fig.~\ref{fig3} (possibly because of our neglect of the anisotropy),
but the fits to $\rho_{xy}$ are very good.

In the inset to Fig.~\ref{fig4}, we show the fitted values of
$\omega_c \tau_H$ (solid line), along with the values of $\omega_c
\tau_{tr}$ (circles) that were used to constrain the fits.  Also included are
the values of $\omega_c \tau_H$ (triangles) obtained from the weak field Hall
angle.  Again, we find that the scattering time determined from the
high field MR agrees better with the $\tau_H$ determined from the weak
field Hall angle than with the $\tau_{tr}$ determined from the
resistivity.

Given the approximations used, we do not claim to favor the particular
two-lifetime model of Eqs.~\ref{cstmr} and~\ref{csthall}.  Rather, the
combination of data from the weak field regime in the optimally doped
sample and the high field regime for the overdoped sample does favor
some kind of two-lifetime model for the cuprate normal state.  It
would be very interesting to analyze high field MR saturation in
cuprates at optimal doping, where separate temperature dependences of
the two scattering times are observed; however, this would require
much higher magnetic fields than are currently available.

In summary, we have succeeded in studying normal state
magnetotransport of single crystal $\rm Tl_2Ba_2CuO_{6+\delta}$
in 60T pulsed
magnetic fields to obtain new constraints on the scattering mechanisms in
the cuprates.  In optimally doped samples, the weak-magnetic-field
regime reaches to fields as high as 60T.  In overdoped samples, we
leave the weak-field regime, as shown by the behavior of both the MR
and the Hall resistance.  Analysis of the MR and Hall effect for both
the optimally doped and overdoped samples gives evidence in support of
those models for the normal state of the cuprates in which two
scattering lifetimes exist at all points on the Fermi surface.


\begin{references}
\bibitem{com1}
Data from the optimally doped sample do show evidence of the high
field regime for $T < 140K$; however, the deviations from parabolicity
are too small to usefully constrain $\beta$ of Eq.~\ref{general}.
Also, interpretation of $T < 130K$ data is complicated by
evidence of superconducting fluctuations.
\bibitem{com2} In some
formulations\cite{carrington,kendziora}, $\rho$ is dominated by
scattering on flat parts of a square FS, while in others\cite{pines}
both the Hall effect and resistivity are determined by FS averages of
``hot'' and ``cold'' scattering rates which have non-integer power law
temperature dependences.
\end{references}
\end{document}